# Does carrier velocity saturation help to enhance $f_{max}$ in graphene field-effect transistors?

Pedro C. Feijoo[1], Francisco Pasadas[1], Marlene Bonmann[2], Muhammad Asad[2], Xinxin Yang[2], Andrey Generalov[3], Andrei Vorobiev[2], Luca Banszerus[4], Christoph Stampfer[4], Martin Otto[5], Daniel Neumaier[5], Jan Stake[2], David Jiménez[1]

1 Universitat Autònoma de Barcelona, 08193 Cerdanyola del Vallès, Spain

2 Chalmers University of Technology, SE-41296 Gothenburg, Sweden

3 Aalto University, FI-00076 Helsinki, Finland

4 2nd Institute of Physics, RWTH Aachen University, 52074 Aachen, Germany

5 Advanced Microelectronic Center Aachen, AMO GmbH, 52074 Aachen, Germany

## Abstract

It has been argued that current saturation in graphene field-effect transistors (GFETs) is needed to get the highest possible maximum oscillation frequency ($f_{max}$). This paper numerically investigates whether velocity saturation can help to get better current saturation and if that correlates with enhanced $f_{max}$. For such a purpose, we used a drift-diffusion simulator that includes several factors that influence output conductance, especially at short channel lengths and/or large drain bias: short-channel electrostatics, saturation velocity, graphene/dielectric interface traps, and self-heating effects. As a testbed for our investigation, we analyzed fabricated GFETs with high extrinsic cutoff frequency $f_{T,x}$ (34 GHz) and $f_{max}$ (37 GHz). Our simulations allow for a microscopic (local) analysis of the channel parameters such as carrier concentration, drift and saturation velocities. For biases far away from the Dirac voltage, where the channel behaves as unipolar, we confirmed that the higher is the drift velocity, as close as possible to the saturation velocity, the greater $f_{max}$ is. However, the largest $f_{max}$ is recorded at biases near the crossover between unipolar and bipolar behavior, where it does not hold that the highest drift velocity maximizes $f_{max}$. In fact, the position and magnitude of the largest $f_{max}$ depend on the complex interplay between the carrier concentration and total velocity which, in turn, are impacted by the self-heating. Importantly, this effect was found to severely limit radio-frequency performance, reducing the maximum $f_{max}$ from ~60 to ~40 GHz.

**Terms:** graphene field-effect transistor, radio-frequency, self-heating, velocity saturation, current saturation, maximum oscillation frequency

## 1. Introduction

The development of new-generation radio-frequency (RF) electronics enables extending the range of advanced applications within the areas of communication, security imaging, quality control, medicine etc. [1]. For the sustainable development, new materials with enhanced electronic properties are required. Graphene is considered as a promising channel material for RF field-effect transistors due to its intrinsically high charge carrier mobility (up to $2 \times 10^5$ cm$^2$ V$^{-1}$ s$^{-1}$) and saturation velocity ($4 \times 10^7$ cm s$^{-1}$) [2–6]. However, RF performance of the graphene field-effect transistors (GFETs) was limited until recently by several factors, for example, a relatively high drain conductance due to zero bandgap, a high graphene/metal contact resistance and the extrinsic carrier scattering by charged defects [7–11]. Continuous efforts in the study of GFETs have resulted in an important improvement of the RF figures of merit: the extrinsic cutoff (transit) frequency ($f_{T,x}$) and maximum frequency of oscillation ($f_{max}$) [12–17]. This enhancement has been enabled, in particular, by the use of GFET models, which have allowed for clarifying and overcoming RF performance limitations [18–27]. Recently, state-of-the-art values of $f_{T,x}$ = 34 GHz and $f_{max}$ = 37 GHz for GFETs with chemical vapor deposited graphene and a gate length ($L_g$) of 500 nm were reported by some of us [28]. These values of $f_{T,x}$ and $f_{max}$ surpassed those of the best Si MOSFETs with similar gate lengths [29]. The achievement has been obtained by a combination of different improvements of the GFET design and fabrication process that have resulted in high saturation velocity, low contact resistance, and reduced extrinsic pad capacitances.

A question that remains to be answered is whether operating the GFET in a saturation velocity regime actually helps to get the highest $f_{max}$. The resolution to this problem needs a simulation tool that considers the factors that affect the current saturation, namely, short-channel effects, velocity saturation effects, and self-heating effects (SHE). Our preliminary analysis indicated that these effects can be significant when a GFET works at relatively high drain fields, above 1 V µm$^{-1}$. In previous works, a self-consistent simulator that accounted for short-channel and velocity saturation effects was developed to investigate the RF performance and scalability of GFETs [24, 25]. The present work updates that simulator by including the SHE and then applies it with two purposes: to study the DC and RF performance of the prototype 500-nm GFET presented in [28] and to explore whether there is still room for $f_{max}$ improvement by exploiting the saturation velocity regime.

To investigate GFET performance, we follow an approach that consists firstly in solving the drift-diffusion equation self-consistently with the two-dimensional Poisson's equation to get the DC characteristics [24]. This set of equations is, in turn, coupled with the heat transfer equation that models the SHE. Then RF performance is obtained from a quasi-static small-signal model, whose parameters are extracted from linearization of the DC simulations [25]. Such a methodology is thoroughly described in Section 2. The combined analysis of DC and RF simulations allows us to assess the influence of graphene electrical properties as the saturation velocity and low-field mobility, and other limiting factors as, for instance, the contact resistance, the interface traps and extrinsic capacitances. Thereby, we have obtained insights on the mechanisms defining the DC and RF performance of GFETs, which are discussed in Section 3. Particularly, we have addressed the question whether velocity saturation can help to get better current saturation and if that correlates with enhanced $f_{max}$. Finally, the conclusions are drawn in Section 4.

## 2. Methods

### 2. 1. Device structure and description of the self-consistent simulator

To investigate the bias dependence of RF performance and its relation with current saturation, we have numerically investigated the prototype GFET with high extrinsic $f_{T,x}$ and $f_{max}$ described in [28]. Figure 1 shows the GFET scanning electron microscope (SEM) image of the device and a schematic view of one of the two fingers. The GFET gate length and total gate width were $L_g$ = 500 nm and $W_g$ = 2 × 15 µm, respectively. The length of each ungated region of the channel was $L_{ung}$ = 100 nm. The graphene layer was encapsulated between insulating layers of $Al_2O_3$ and $SiO_2$ with thicknesses of $t_t$ = 22 nm and $t_b$ = 1 µm, respectively. The relatively thick $SiO_2$ allows for reduction of the parasitic pad capacitances [28]. High-resisitivy (larger than 10 kΩ cm) silicon was used as substrate with the aim of minimizing the substrate-related microwave loss in the GFET contact pads and transmission lines of the prospective devices [30]. Details on the GFET fabrication are included in section S1 of the supplementary information.

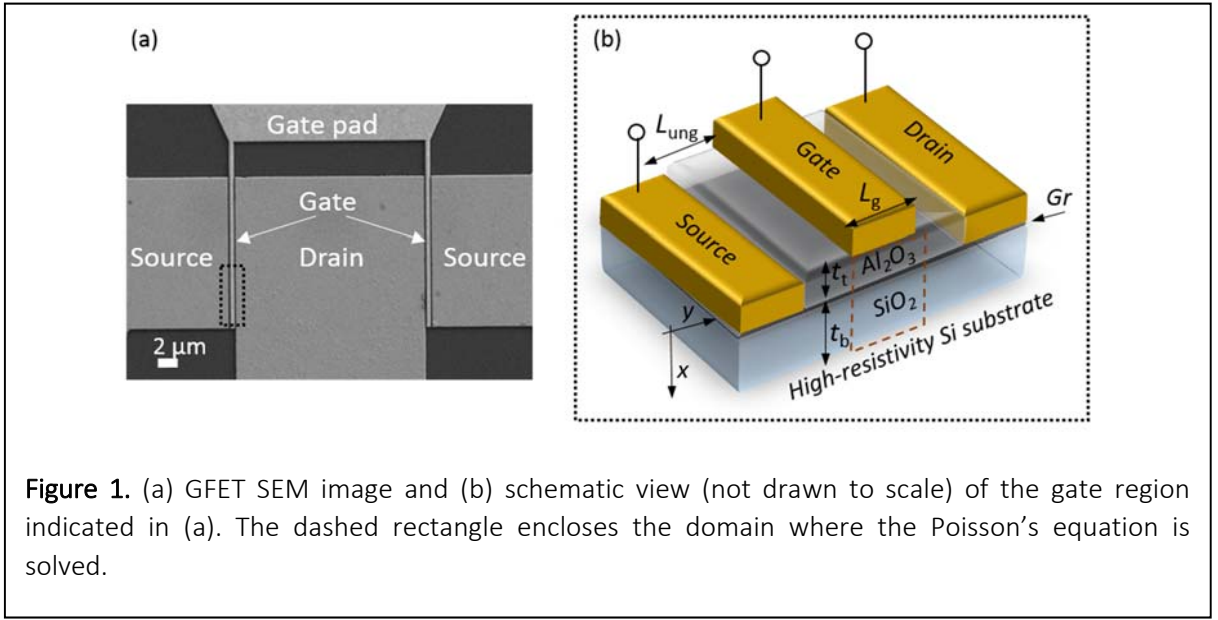

**Figure 1.** (a) GFET SEM image and (b) schematic view (not drawn to scale) of the gate region indicated in (a). The dashed rectangle encloses the domain where the Poisson's equation is solved.

The GFET was simulated using the method described in [24, 25], which consists in solving self-consistently 2D Poisson's equation and 1D drift-diffusion transport equation. The dashed rectangle in figure 1(b) encloses the active area of the transistor and corresponds to the domain where the Poisson's equation is solved. The simulator obtains the stationary distributions of graphene electrical parameters along the channel as a function of the voltages applied to the gate-source and drain-source terminals ($V_{gs}$ and $V_{ds}$, respectively). Specifically, it is possible to get the local parameters such as the charge carrier concentration for both electrons ($n$) and holes ($p$), the carrier field-dependent mobility ($\mu$), the separate currents and carrier velocities driven by both drift ($v_{drift}$) and diffusion mechanisms ($v_{diff}$), the Dirac energy ($E_D$=-$q\psi$) and the quasi-fermi energy ($E_F$=-$qV$). The details of the simulator and the different carrier velocity definitions used in this work can be found in sections S2 and S3, respectively, of the supplementary information. Key parameters as the flatband voltage ($V_{gs0}$), the residual charge carrier concentration ($\rho_0$), the low-field mobility ($\mu_{LF}$), and the contact resistance ($R_c$) were extracted from measured low-$V_{ds}$ transfer curves ($I_{ds}$-$V_{gs}$) with holding time of 1 s at each bias point (see section S4 of the supplementary information). After that, we fitted the measured output characteristics ($I_{ds}$-$V_{ds}$), which were obtained upon application of a holding time of 30 s per measured point. That time is long enough for the trapping/de-trapping processes to stabilize at high fields [31]. The fitting parameters are the interface trap density ($N_{it}$), the energy of optical phonons ($\hbar\Omega$), whose emission limits carrier drift velocity, and the thermal resistance ($R_{th}$). The latter will be discussed below. The model for the saturation velocity $v_{sat}$ is given by [32]:

$$v_{sat}(y) = \frac{2\Omega}{\pi\sqrt{\pi\rho_{sh}(y)}} \sqrt{1 - \frac{\Omega^2}{4\pi v_F^2 \rho_{sh}(y)}} \frac{1}{N_{OP}+1} \tag{1}$$

where $T$ is the temperature, and $\rho_{sh}(y) = n(y) + p(y)$ is the local carrier concentration at the position $y$ in the channel. The phonon occupation $N_{OP}$ depends on temperature as:

$$N_{OP} = \frac{1}{\exp\left(\frac{\hbar\Omega}{kT}\right) - 1} \tag{2}$$

Eqs. (1) and (2) show that saturation velocity strongly depends on carrier concentration. Moreover, an increase in temperature slightly decreases $v_{sat}$. These dependencies can be seen in figure 2, where $v_{sat}$ has been represented for typical values of carrier concentration at several temperatures.

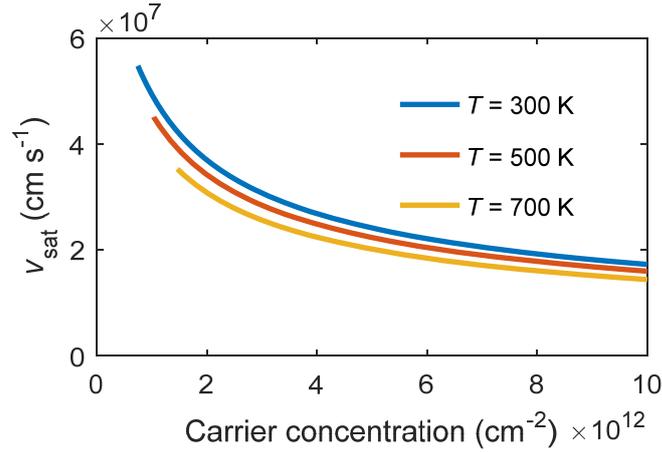

**Figure 2.** Carrier saturation velocity as a function of the carrier concentration for different temperatures. The value of the optical phonon energy used has been $\hbar\Omega = 0.10$ eV.

We have used the following equation to model the field-dependent mobility $\mu(y)$ as a function of the local electric field and saturation velocity, and thereby, also on $\rho_{sh}(y)$ and $T$:

$$\mu(y) = \frac{\mu_{\text{LF}}}{\left\{ 1 + \left[ \left( \frac{\mu_{\text{LF}}}{v_{\text{sat}}(y)} \left| \frac{\partial \psi}{\partial y} \right| \right)^{\gamma} \right] \right\}^{\frac{1}{\gamma}}} \tag{3}$$

Here, a value of 1 has been used for the parameter $\gamma$, consistently with numerical studies of electronic transport in single layer graphene relying on Monte Carlo simulations [33].

Unlike our previous works, we have included the SHE in the self-consistent loop of the GFET simulator. This means that we assume that the temperature of the GFET rises because the heat dissipated in graphene by the Joule effect finds difficulty to spread out of the device through the surrounding layers. By using the simple equivalent thermal circuit of figure 3(a) to describe the SHE, the temperature of the graphene channel can be expressed as:

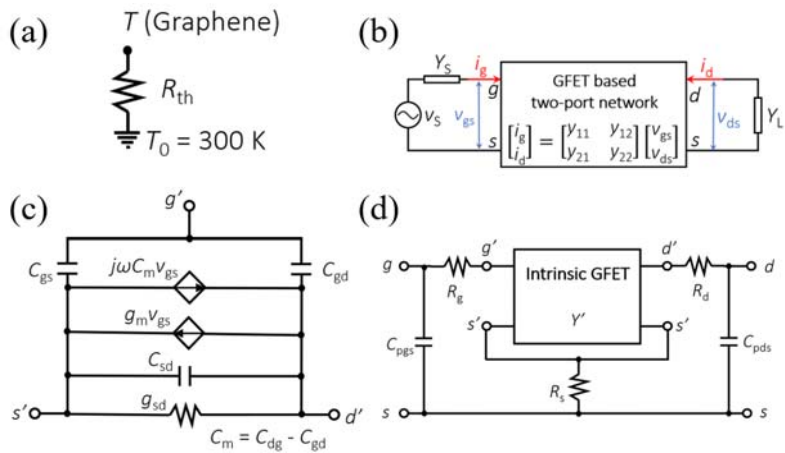

**Figure 3.** (a) Thermal model assumed for the GFET. (b) GFET configured as a two-port network, with the input port formed by the gate-source terminals and the output port by the drain-source terminals. (c) Intrinsic equivalent circuit of the GFET. (d) Extrinsic embedding network with the parasitic series resistances, $R_g$, $R_d$, and $R_s$ and the parasitic capacitances between gate and source ($C_{pgs}$) and between drain and source ($C_{pds}$).

$$T - T_0 = R_{\text{th}} P_{\text{dis}} \qquad (4)$$

where $T_0$ = 300 K is the temperature of the heat sink, assumed to be the substrate, and $P_{\text{dis}}$ is the dissipated power in the GFET, which takes the following form:

$$P_{\text{dis}} = |I_{\text{ds}} V'_{\text{ds}}| \qquad (5)$$

where $V'_{\text{ds}} = V_{\text{ds}} - I_{\text{ds}} R_c$ is the intrinsic drain-to-source voltage. This model considers an average temperature for the whole graphene sheet, so it neglects any local temperature deviation.

Using the values for mobility and carrier concentration obtained in this study we estimate the mean free path (MFP) by the semiclassical model described in [6] in the 10-100 nm range. Since the MFP is much shorter than the source-to-drain length ($L_g + 2L_{\text{ung}}$), it is confirmed that the drift-diffusion transport mechanism is appropriate for describing the electronic transport in the examined GFET.

## 2.2. Small-signal model of the GFET and derived RF performance

For the analysis of the RF performance, we consider the GFET as a two-port network in common-source configuration as depicted in figure 3(b). The device is connected to source and load admittances and is characterized by its extrinsic admittance matrix $Y$. This matrix is calculated in two steps. First, the intrinsic admittance matrix $Y'$ is determined by the intrinsic small-signal equivalent circuit model of figure 3(c) assuming quasi-static operation [34]. Then the extrinsic $Y$ matrix is obtained embedding the intrinsic GFET in a simplified extrinsic circuit of lumped elements, which consists of parasitic resistances at each of the three terminals and parasitic capacitances at both the input and the output ports, as shown in figure 3(d). Since $t_b \gg t_t$, the back-gate capacitance is much smaller than the top-gate capacitance, so we can neglect the influence of the substrate capacitance in $Y'$. Tunneling currents through any of the dielectrics are also neglected.

Transconductance $g_m$ and output conductance $g_{\text{sd}}$ can be obtained from the derivatives of $I_{\text{ds}}$ respect to the intrinsic bias voltages $V'_{\text{gs}} = V_{\text{gs}} - I_{\text{ds}} R_c/2$ and $V'_{\text{ds}}$, respectively. Then, the small-signal capacitances are determined from the charges associated to each of the terminals ($Q_i$, with $i$ = s, d or g). They have been defined assuming a charge conserving Ward-Dutton's linear charge partition scheme [34]. The transcapacitances $C_{\text{gs}}$, $C_{\text{gd}}$, $C_{\text{sd}}$, $C_{\text{dg}}$ are obtained as the derivative of charge at terminal $i$ with respect to the intrinsic voltage at terminal $j$, $C_{ij} = -dQ_i/dV'_j$. For the calculation of the small-signal parameters, we assume that the temperature is constant at a given bias point. A full description of the small-signal parameter calculation can be found in section S5 of the supplementary information. Finally, the RF figures of merit $f_{T,x}$ and $f_{\text{max}}$ are extracted from the current gain and unilateral power gain that result from the $Y$ matrix [35].

**Table 1.** Optimized parameters fitted from current-voltage characteristics.

| Parameter | Optimized value |
|---|---|
| $V_{\text{gs0}}$ | 2.2 V |
| $\rho_0$ | $2.9 \cdot 10^{11}$ cm$^{-2}$ |
| $\mu_{\text{LF}}$ | $2.0 \cdot 10^3$ cm$^2$ V$^{-1}$ s$^{-1}$ |
| $R_c$ | 11 Ω |
| $N_{\text{it}}$ | $<10^{12}$ eV$^{-1}$ cm$^{-2}$ |
| $\hbar\Omega$ | 0.10 eV |
| $R_{\text{th}}$ | $2.7 \cdot 10^4$ K W$^{-1}$ |

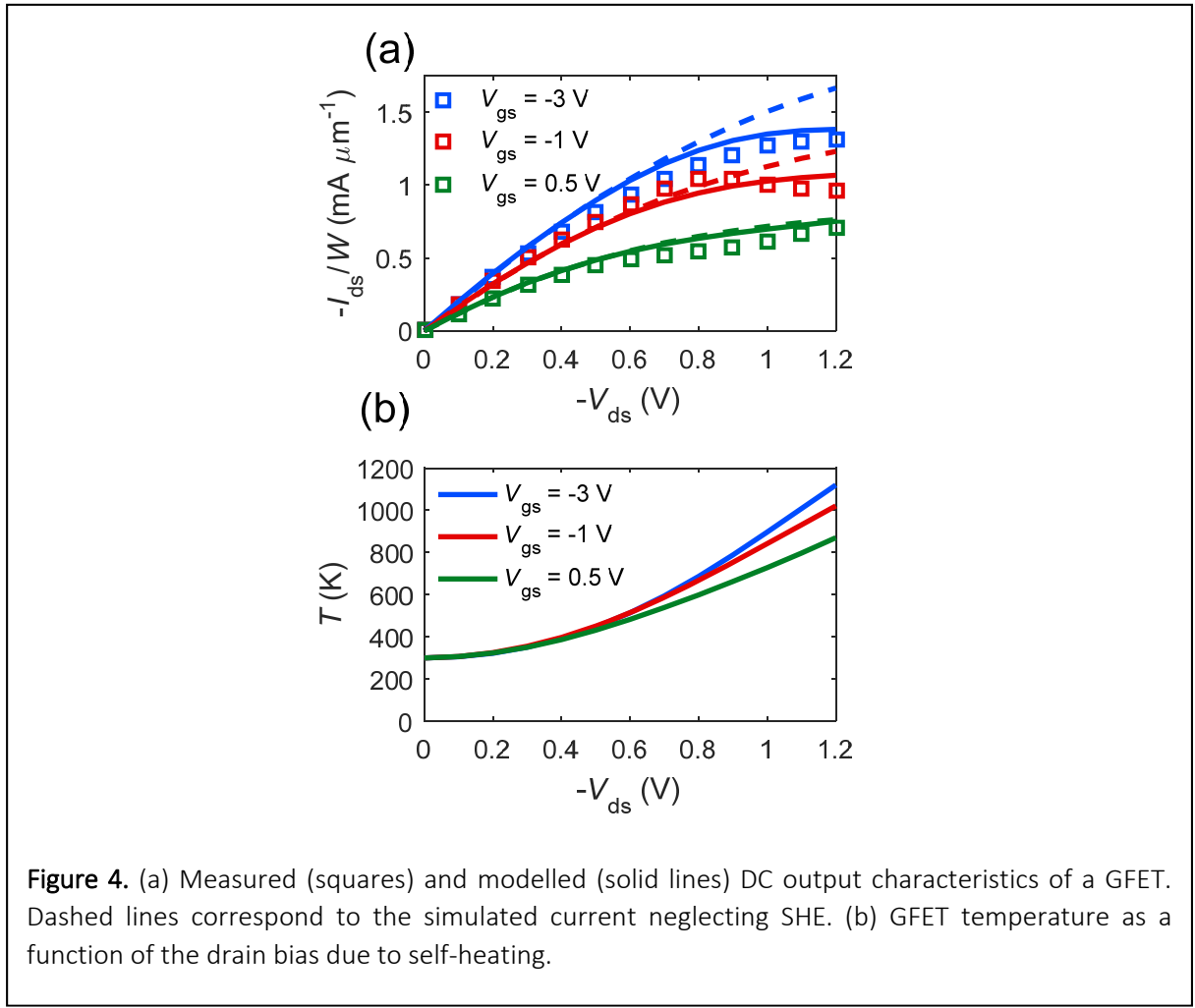

**Figure 4.** (a) Measured (squares) and modelled (solid lines) DC output characteristics of a GFET. Dashed lines correspond to the simulated current neglecting SHE. (b) GFET temperature as a function of the drain bias due to self-heating.

## 3. Results and discussions

First, we reproduced the experimental DC characteristics following the methodology described in Section 2.1. Figure 4(a) shows the measured output characteristics of the GFET and their comparison with the simulations, where the parameters used are presented in Table 1. The estimated $N_{it}$ was found to be lower than $10^{12}$ eV$^{-1}$ cm$^{-2}$, a value below which the influence of interface traps is negligible, as shown in section S6 of the supplementary information. The best fitting value of the $R_c$ is 11 $\Omega$, which includes the metal/graphene contact resistance at both drain and source together with the access resistances of the ungated graphene regions. Thus, the width specific contact resistivity results in $R_c \cdot W_g/2 = 165$ $\Omega$ $\mu$m. The width specific metal/graphene contact resistivity of approximately 90 $\Omega$ $\mu$m, evaluated applying the drain resistance model to the transfer characteristic, agrees with the value of 95 $\Omega$ $\mu$m obtained by transfer length measurements (which exclude access resistance). From the fitting done in figure 4(a), we obtained a value of $2.7 \cdot 10^4$ K W$^{-1}$ for the $R_{th}$ shown in figure 3(a), which agrees with the order of magnitude of calculations based on the model by Pop *et al.* [22, 32], of around $3 - 4 \cdot 10^4$ K W$^{-1}$. Our simulator also allowed us to calculate GFET temperature as a function of the bias, shown in figure 4(b), and ranging between 300 and 1000 K.

Next, we have benchmarked the small-signal model against the experimental $Y$-parameters. Figure 5 shows the measured $Y$-parameters in the 1-50 GHz range at $V_{ds}$ =-1.1 V and $V_{gs}$ = 0.5 V, which correspond to the bias with the highest measured $f_{T,x}$ =34 GHz and $f_{max}$ =37 GHz. The four elements of the complex admittance matrix are compared against calculations obtained using the equivalent circuit of figure 3(d). The intrinsic $Y'$ was directly extracted from the quasi-static small-signal model given in figure 3(c), while the values of gate series resistance $R_g$, the parasitic capacitances $C_{pgs}$ and $C_{pds}$ were optimized to fit the measured $Y$-parameters. Both series resistances at drain and source, $R_d$ and $R_s$, were assumed to be $R_c/2$. In addition to the good agreement between simulated and measured $Y$-parameters in the whole range of examined frequencies, figure 6 shows that the extracted values of $C_{pgs}$ and $C_{pds}$, presented in Table 2, are similar to the ones measured from an open GFET structure (i.e. without the graphene layer), which confirms the validity of our approach.

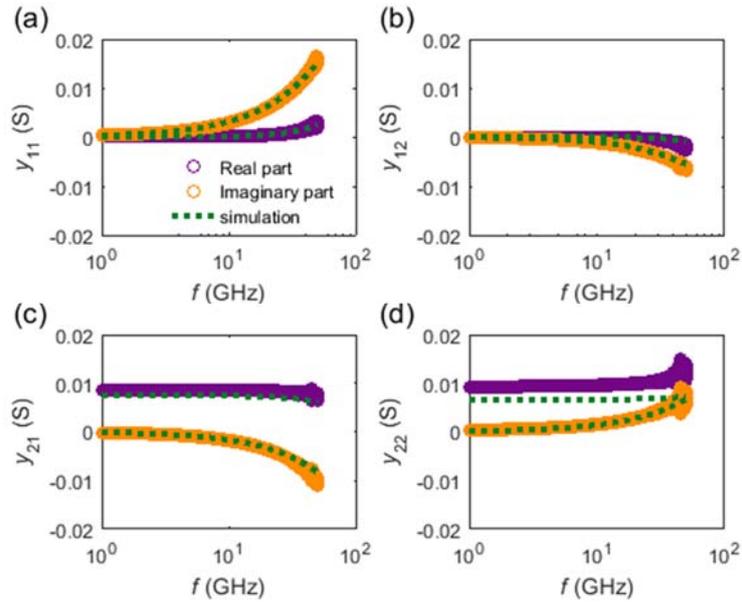

**Figure 5.** Measured (circles) and modelled (dotted lines) Y-parameters as a function of the frequency for a GFET biased at $V_{ds}$ = -1.1 V and $V_{gs}$ = 0.5 V, corresponding to the highest measured $f_{T,x}$ = 34 GHz and $f_{max}$ = 37 GHz.

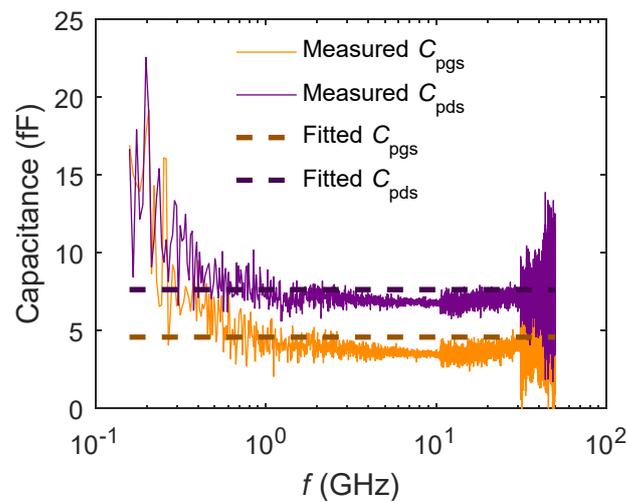

**Figure 6.** Parasitic capacitances measured from the open structure of the GFET and their values obtained from $Y$-parameters fitting.

**Table 2.** Values of the parasitic elements in the GFET extrinsic network.

| Parameter | Optimized value |
|-----------|-----------------|
| $R_g$ | 10 Ω |
| $R_s$ | 5.5 Ω |
| $R_d$ | 5.5 Ω |
| $C_{pgs}$ | 4.6 fF |
| $C_{pgs}$ | 7.6 fF |

Using the parasitic elements found in the previous step, we analyzed the bias dependence of $f_{T,x}$ and $f_{max}$. The results are compared with measurements in figure 7, showing similar trends. A more detailed insight on the bias dependence of RF performance can be obtained from the map of $f_{max}$ shown in figure 8(a). A total of four maxima with $f_{max}$ of ~40 GHz are observed and labelled as A, B, C, and D, where A and C maxima occur at positive drain bias while B and D maxima at negative drain bias. Note that when gate voltage is equal to the Dirac voltage (*i. e.* $V_{gs} = V_D \approx V_{gs0} + V_{ds}/2$), transconductance $g_m$ changes its sign, which makes $f_{max} \sim 0$. On top of that, for a given drain bias polarity, *e. g.* negative, the B maximum is located at $V_{gs} < V_D$, which corresponds to a unipolar p-channel with the pinch-off close to the drain side, while at $V_{gs} > V_D$, the D maximum corresponds to a unipolar n-channel with pinch-off close to the source side (see carrier distributions shown in figure 9). Those maxima A, B, C and D are located at biases where there is a drop in the total carrier concentration close to the source or to the drain edges.

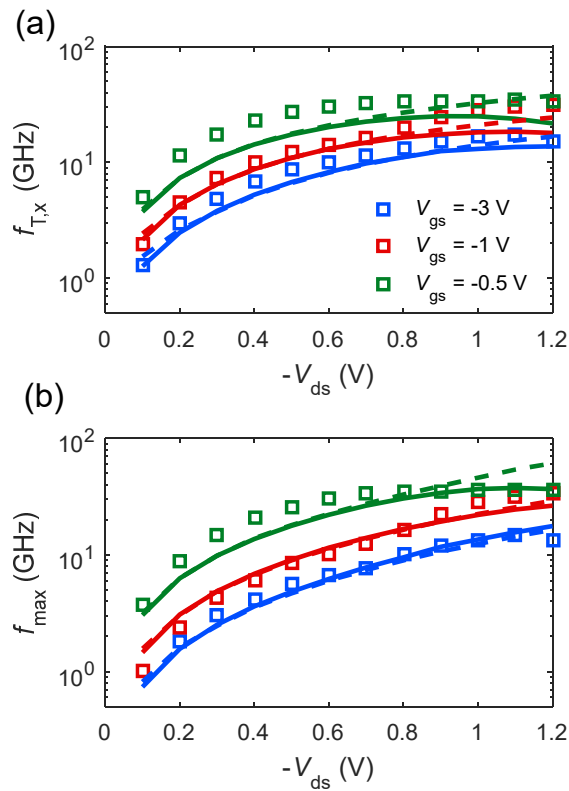

**Figure 7.** Measured (squares) and modelled (solid lines) of (a) $f_{T,x}$ and (b) $f_{max}$ as a function of the drain bias. Dashed lines represent figures of merit $f_{max}$ and $f_{T,x}$ switching off the self-heating effect.

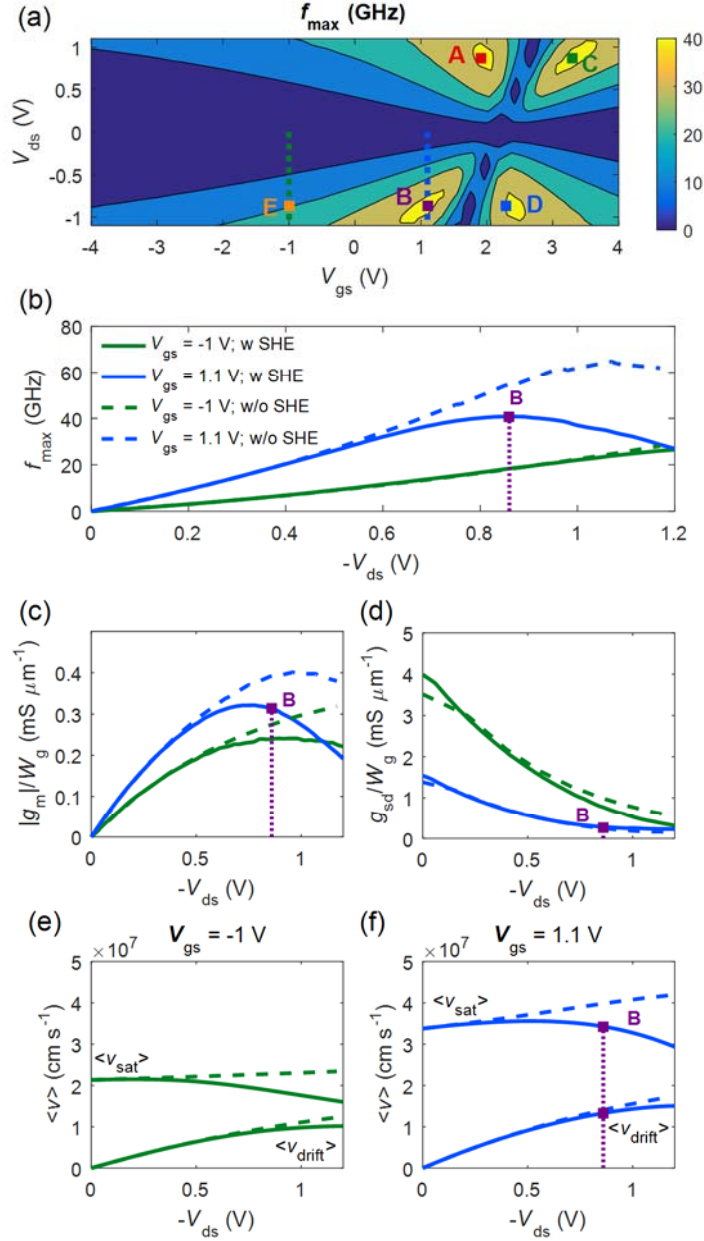

**Figure 8.** (a) Bias dependent $f_{max}$. It presents four maxima labelled as A, B, C and D. Dashed lines represent the locus of considered drain bias points for $f_{max}$ plot shown in (b). Transconductance, $g_m$, and output conductance, $g_{sd}$, are represented in (c) and (d), respectively. (e) and (f) show average drift velocity and saturation velocity, respectively. Dashed lines correspond to the case where SHE have been switched off.

It has been argued that the highest $f_{max}$ needs current saturation in GFETs. To get the desired current saturation, it has been proposed that GFET operation close to the carrier velocity saturation regime is helpful. Here we critically review this idea by comparison of the bias-dependent $f_{max}$ and $g_m$ maps plotted in figures 8(a) and 9(c), respectively. First observation is that the bias location of the four $f_{max}$ maxima roughly coincide with the $|g_m|$ maxima. For a deeper insight in figure 8(a), we analyzed $f_{max}$ evolution at two different constant $V_{gs}$, the first at $V_{gs} = 1.1$ V passing through the maximum B (dashed blue line), and the second at $V_{gs} = -1.0$ V (dashed green line) passing far away from the maximum B.

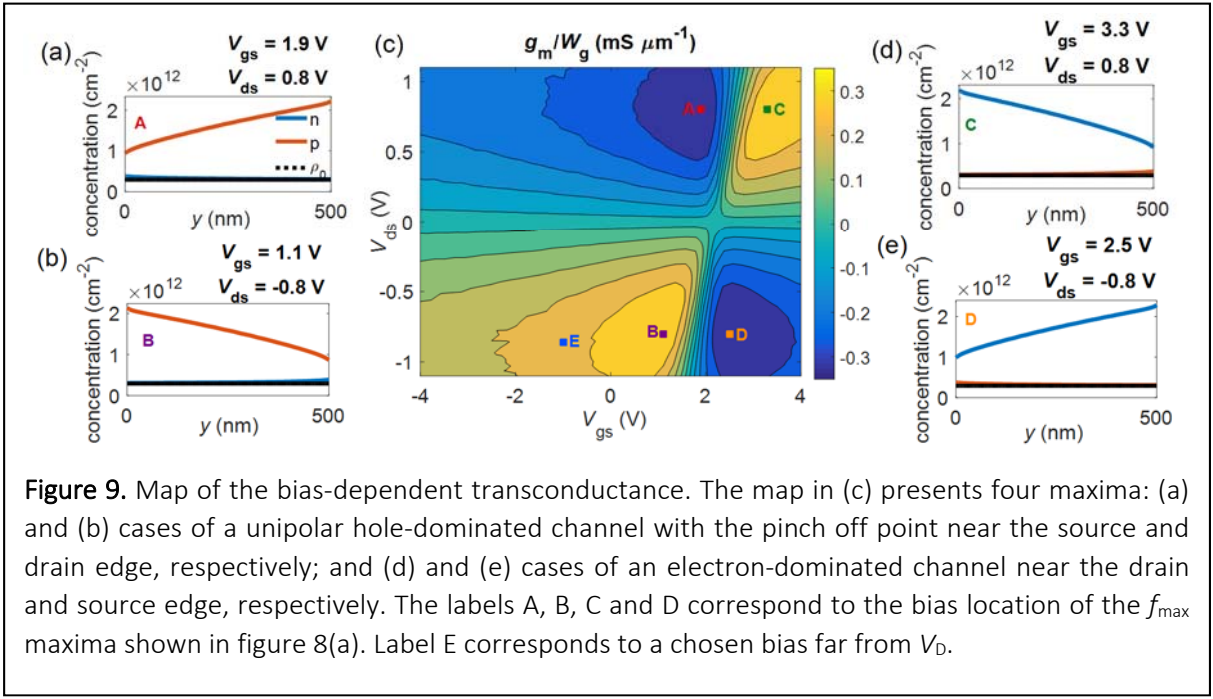

**Figure 9.** Map of the bias-dependent transconductance. The map in (c) presents four maxima: (a) and (b) cases of a unipolar hole-dominated channel with the pinch off point near the source and drain edge, respectively; and (d) and (e) cases of an electron-dominated channel near the drain and source edge, respectively. The labels A, B, C and D correspond to the bias location of the $f_{max}$ maxima shown in figure 8(a). Label E corresponds to a chosen bias far from $V_D$.

The resulting plot is shown in figure 8(b). For $V_{gs}$ = 1.1 V case the B maximum is reached at $V_{ds}$ = -0.86 V, while for $V_{gs}$ = -1.0 V there is no an absolute maximum of $f_{max}$, being $f_{max}$ a monotonous function of $V_{ds}$, instead. Analyzing the average drift velocity in Figure 8(e), we confirm the expectation that, far from the B maximum, the higher the drift velocity is (even approaching the saturation velocity), the better $f_{max}$ and the current saturation are, as shown in figures 8(b) and (d), respectively. This behavior indeed happens for biases far away from the Dirac voltage, where the channel behaves as unipolar. However, the largest $f_{max}$ is observed at biases near the crossover between unipolar and bipolar behavior such as the B point, where it does not hold that the highest drift velocity, represented in figure 8(f), gives the largest $f_{max}$, represented in figure 8(b). In fact, figure 8(d) shows that $g_{sd}$ does not reach a minimum value at $V_{ds}$ = -0.86 V. Instead, highest $f_{max}$ is reached close to the maximum of $|g_m|$, as shown in figure 8(c). The bias point B and the bias corresponding to the maximum of $|g_m|$ slightly differ because $f_{max}$ depends in a complex way not only on $g_m$, but on $g_{sd}$, the transcapacitances and the parasitic elements [36]. On the other hand, $f_{max}$ is not the highest possible when the GFET is operated far away from Dirac voltage, and this can be explained by the degraded $g_m$ and $g_{sd}$, as shown in figures 8(c) and 8(d), respectively. The degradation of $g_m$ and $g_{sd}$ at bias E respect to the bias B is caused, in turn, by a decrease in the drift velocity because of the larger carrier concentration. The bias that maximizes RF performance is thus the result of a complex interplay between carrier concentration and carrier velocity in graphene, where self-heating plays a significant role.

A local analysis of the carrier velocities along the channel at both biases E (figure 10) and B (figure 11) reveals more details on the central question of this paper, namely, if velocity saturation is needed for the highest $f_{max}$. As there are two transport mechanisms at play (drift and diffusion), we have introduced in S3 of the supplementary information, as a matter of convenience, the definitions of drift, diffusion and total velocities that can be directly compared with the saturation velocity. At the E bias, where the channel is unipolar p-type, figure 10(b) shows that $v_{drift}$ dominates over $v_{diff}$ and is roughly 50% of $v_{sat}$. The ratio $v_{drift}/v_{sat}$ could be increased up to ~100 % with a higher drain bias; for instance, it is 64% for $V_{ds}$=-1.2 V, according to figure 8(e). However, at the B point, where the pinch-off is near the drain side, diffusion contribution is much higher with $v_{diff}/v_{drift}$ around 40% near the drain, being $v_{drift}/v_{sat}$ ~ 45 %, as can be seen in figure 11(a). Therefore, our results do not support that operating in the regime of velocity saturation results in the highest $f_{max}$.

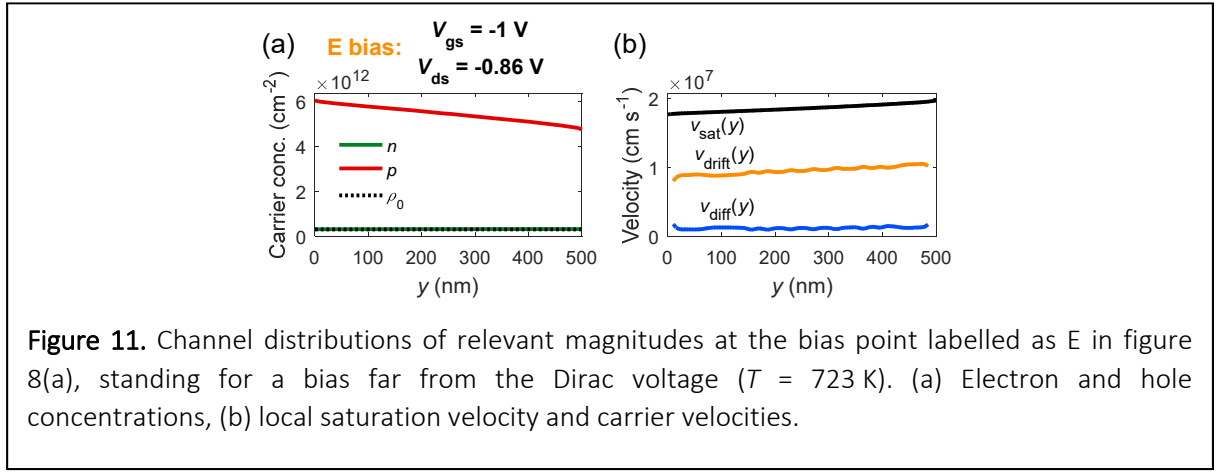

**Figure 11.** Channel distributions of relevant magnitudes at the bias point labelled as E in figure 8(a), standing for a bias far from the Dirac voltage ($T$ = 723 K). (a) Electron and hole concentrations, (b) local saturation velocity and carrier velocities.

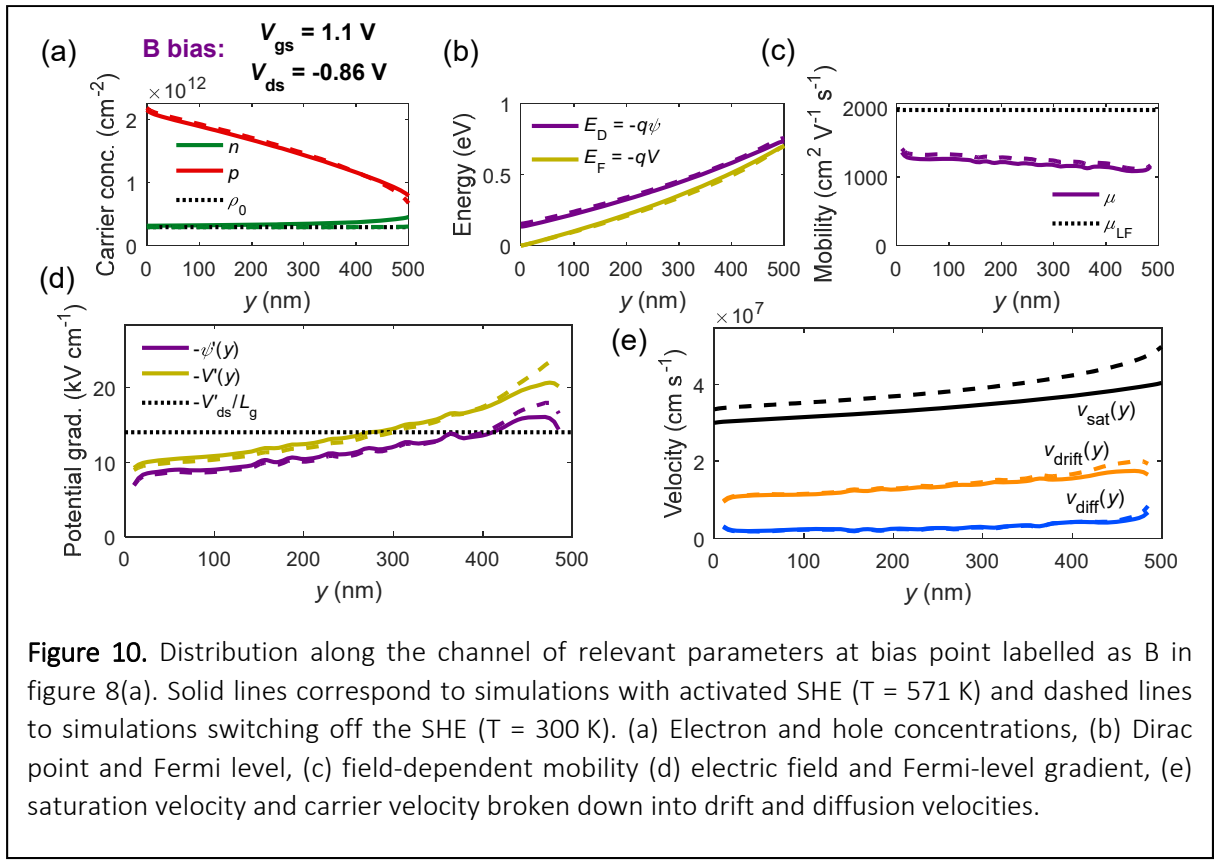

**Figure 10.** Distribution along the channel of relevant parameters at bias point labelled as B in figure 8(a). Solid lines correspond to simulations with activated SHE (T = 571 K) and dashed lines to simulations switching off the SHE (T = 300 K). (a) Electron and hole concentrations, (b) Dirac point and Fermi level, (c) field-dependent mobility (d) electric field and Fermi-level gradient, (e) saturation velocity and carrier velocity broken down into drift and diffusion velocities.

To assess the impact of SHE, we have shown in figure 4(a) how SHE affects drain current, after switching it on and off in the simulations. It can be observed that current saturation is a result of self-heating, which is triggered at $|V_{ds}| > 0.6$ V. Additionally, figures 7, 8 and 11 also show the SHE impact on the different parameters of the GFET. At biases near maximum values of $f_{max}$, SHE are prominent and graphene temperature reaches $\sim 700$ K. Importantly, the value of $f_{max}$ can be degraded from 60 to 40 GHz mainly due to a decrease in $v_{drift}$, which reduces $g_m$ from 0.4 to 0.3 mS $\mu m^{-1}$ despite the larger $v_{drift}/v_{sat}$ ratio. This way, it can be concluded that high temperatures limit the RF performance of GFETs. More details on the effect of SHE on RF performance can be found in section S7 of the supplementary information.

## 4. Conclusions

In this work we analyzed the influence of carrier velocity saturation on the RF performance by means of a drift-diffusion self-consistent GFET simulator. The model includes a number of effects defining the current saturation, namely, the two-dimensional electrostatics, saturation velocity effects, and self-heating effects, which are especially relevant at short channels and/or large drain bias. First, we optimized the model parameters to fit the experimental DC characteristics of a prototype GFET. Then, the measured *Y* parameters were reproduced by fitting the values of the parasitic capacitances and the gate series resistance and simulated GFET small-signal equivalent circuit. We simulated the bias dependence of the measured RF figures of merit and we discussed the role played by saturation velocity in defining the highest $f_{max}$. For biases far from the Dirac voltage, where the channel is unipolar, a higher drift velocity results in a larger $f_{max}$. However, the largest $f_{max}$ are located at biases close to the onset of bipolar conduction. In that scenario the pinch-off point is close to the source or drain edge and drift velocity there is no longer saturated. This is caused by the combined effects of carrier concentration and total velocity, which are interdependent in a complex way. Notably, we found a significant degradation of $f_{max}$ at high drain bias because of the self-heating. Based on these results, further optimization of the GFET design for applications in advanced RF electronics can be done by selecting the appropriate bias and reducing the effects of self-heating.

## Acknowledgments

This work was funded in part by the European Union's Horizon 2020 Research and Innovation Program under Grant GrapheneCore2 785219, in part by the *Generalitat de Catalunya* under Grant 2017 SGR 954, in part by the *Ministerio de Economía y Competitividad* under Grant TEC2015-67462-C2-1-R, in part in part by the Swedish Research Council under Grant No 2017-04504, and in part by the Academy of Finland.

Supplementary information

# Does carrier velocity saturation help to enhance $f_{max}$ in graphene field-effect transistors?

Pedro C. Feijoo[1], Francisco Pasadas[1], Marlene Bonmann[2], Muhammad Asad[2], Xinxin Yang[2], Andrey Generalov[3], Andrei Vorobiev[2], Luca Banszerus[4], Christoph Stampfer[4], Martin Otto[5], Daniel Neumaier[5], Jan Stake[2], David Jiménez[1]

[1] Universitat Autònoma de Barcelona, 08193 Cerdanyola del Vallès, Spain
[2] Chalmers University of Technology, SE-41296 Gothenburg, Sweden
[3] Aalto University, FI-00076 Helsinki, Finland
[4] 2nd Institute of Physics, RWTH Aachen University, 52074 Aachen, Germany
[5] Advanced Microelectronic Center Aachen, AMO GmbH, 52074 Aachen, Germany

## S1. Fabrication and characterization of GFETs

The two-finger gate GFET analyzed in this work was fabricated as described in [1]. The reported state-of-the-art RF performance is achieved by a combination of improvements of the GFET design and fabrication process. First, a very high-quality CVD graphene film, with a Hall mobility up to 7000 cm$^2$ V$^{-1}$ s$^{-1}$, was transferred to a high-resistivity Si/SiO$_2$ substrate with an increased SiO$_2$ thickness of 1 μm, which resulted in a reduction in the parasitic pad capacitances. Immediately after the transfer, the graphene film was covered by a 5 nm thick protective Al$_2$O$_3$ layer. The protective layer encapsulates graphene and prevents it from contamination during further processing, thereby reducing concentration of traps and charged scattering impurities at the graphene/dielectric interface. Apparently, the use of a buffered oxide etching for opening contact windows in the protective Al$_2$O$_3$ layer resulted in a more effective removal of e-beam resist and PMMA residues and, hence, an extremely low graphene/metal specific contact resistivity, as low as 90 Ω μm.

The DC current-voltage curves and the scattering parameters $S$ of the GFETs were measured using a Keithley 2612B dual-channel source meter and an Agilent N5230A network analyzer, respectively. The RF measurement setup was calibrated at the ground-source-ground microwave probe tips using a CS-5 calibration substrate. The output characteristics ($I_{ds}$-$V_{ds}$) were obtained during the $S$-parameter measurements with a holding time of 30 s at each bias point. This holding time is long enough for the trapping/detrapping processes to stabilize. The $S$-parameters were measured under different bias conditions in the frequency range of 1–50 GHz and used to calculate the admittance parameters $Y$, the small-signal current gain ($h_{21}$) and the unilateral power gain ($U$) [2–4]. The experimental values of $f_{T,x}$ and $f_{max}$ were found as the frequencies at which the magnitudes of $|h_{21}|$ and $U$ equals to 0 dB.

## S2. Self-consistent simulator including self-heating effect

Here we explain the fundamentals of the self-consistent simulator used to calculate the behavior of GFETs with a structure as depicted in figure 1(b) of the main text. More details of the simulator can be found thoroughly described in [5]. The intrinsic bias voltages applied to the electrodes of top gate and drain with respect to the source ($V'_{gs}$ and $V'_{ds}$, respectively) induce a sheet charge density $\sigma(y) = q[p(y) - n(y)] + \sigma_{it}(y)$ in the graphene layer. The magnitudes $p(y)$ and $n(y)$ are the hole and electron concentrations along the graphene channel, $q$ is the elementary charge, $y$ is the axis that goes from source ($y = 0$) to drain ($y = L_g$), where $L_g$ is the channel length. Here $\sigma_{it}(y)$ corresponds to the interface trapped charge density. The sheet charge distribution is needed to calculate the electrostatic potential $\psi(x,y)$ inside the GFET by means of the Poisson's equation. Figure S1 shows the two-dimensional domain where this equation is solved, where $x$ is the position along the axis that goes from back to top gate electrodes. Assuming that the GFET width $W_g$ (in the $z$ direction) is large as compared with the other dimensions of the device, the Poisson's equation can be written as follows:

$$\nabla \cdot [\varepsilon_r(x,y)\varepsilon_0 \nabla \psi(x,y)] = \rho_{free}(x,y) \tag{S1}$$

where $\varepsilon_0$ is the vacuum dielectric constant, and $\varepsilon_r(x,y)$ is the relative dielectric constant, which is equal to $\varepsilon_t$ and $\varepsilon_b$ inside the top and back dielectrics, respectively, and $\varepsilon_G$ in the graphene. From figure S1, the parameters $t_t$ and $t_b$ correspond to the top and back insulator thicknesses, respectively. The charge density $\rho_{free}(x,y)$ is zero inside both dielectrics so its only contribution corresponds to $\sigma(y)$ inside graphene. When solving the Poisson's equation, the electrostatic potential on the top gate is set to $V'_{gs} - V_{gs0}$ and the back gate to $V'_{bs} - V_{bs0}$, where $V_{gs0}$ and $V_{bs0}$ are the flatband voltages. Homogeneous Neumann's conditions are applied to the other two boundaries of the dielectrics to ensure charge neutrality.

The drift-diffusion equation for the drain current $I_{ds}$ reads as follows:

$$I_{ds} = qW_g[n(y) + p(y)]\mu(y)\frac{dV(y)}{dy} \tag{S2}$$

where $\mu(y)$ is the field-dependent mobility, assumed to be equal for electrons and holes, and $V(y)$ is the quasi-Fermi potential in the graphene. The boundary conditions make $V(y)$ equal to zero at $y = 0$ and equal to $V'_{ds}$ at $y = L_g$. Electron and holes share the same quasi-Fermi level due to a very short recombination time of carriers in graphene, of around 10 - 100 ns [6, 7].

In this work we have included the effect of self-heating when calculating charges and current at a certain DC bias. This means that the temperature of the GFET rises due to the heat dissipated by the current flow along the graphene channel and is not properly removed from the device because of the thermal resistance of the surrounding layers. We thus solve self-consistently the previous two-equation system (drift-diffusion and Poisson's equations) together with the solution of the equivalent thermal circuit of figure 3(d) in the main text. The temperature of the graphene channel $T$ must increase as:

$$T - T_0 = R_{\text{th}}P_{\text{dis}} \tag{S3}$$

where $T_0 = 300$ K is the temperature of the heat sink, assumed to be the substrate, and $P_{\text{dis}}$ is the dissipated power in the GFET. In this work, the value of thermal resistance $R_{\text{th}}$ has been considered as a fitting parameter to reproduce the experimental current-voltage curves. $P_{\text{dis}}$ takes the form:

$$P_{\text{dis}} = |I_{\text{ds}}V'_{\text{ds}}| \tag{S4}$$

From both the electrostatic and quasi-Fermi potentials, the carrier concentrations are calculated using the linear dispersion relation of graphene, and thus accounting for its quantum capacitance:

$$n(y) = \rho_0 + N_{\text{G}}\mathcal{F}_1\left[q\,\frac{\psi(0,y)-V(y)}{kT}\right] \tag{S5a}$$

$$p(y) = \rho_0 + N_{\text{G}}\mathcal{F}_1\left[-q\,\frac{\psi(0,y)-V(y)}{kT}\right] \tag{S5b}$$

$$\sigma_{\text{it}}(y) = -q^2 N_{\text{it}}[\psi(0,y)-V(y)] \tag{S5c}$$

We have added the contribution of graphene puddles $\rho_0$ to the carrier concentrations [8]. Here, $k$ is the Boltzmann constant, $T$ is the temperature, $N_{\text{it}}$ is the density of defects, which is assumed to be constant, and $N_{\text{G}}$ is the effective density of states of graphene, given by:

$$N_{\text{G}} = \frac{2}{\pi}\left(\frac{kT}{\hbar v_{\text{F}}}\right)^2 \tag{S6}$$

being $\hbar$ the reduced Planck's constant and $v_{\text{F}}$ the Fermi velocity ($10^8$ cm s$^{-1}$). In equation (S5), $F_1(z)$ refers to the first order Fermi-Dirac integral:

$$\mathcal{F}_i(z) = \frac{1}{\Gamma(i+1)}\int_0^\infty \frac{u^i \mathrm{d}u}{1+e^{u-z}} \tag{S7}$$

The field-dependent mobility model that we have used in this work includes velocity saturation effects in the following form:

$$\mu(y) = \frac{\mu_{\text{LF}}}{\left\{1+\left[\frac{\mu_{\text{LF}}}{v_{\text{sat}}(y)}\left|\frac{\partial\psi(0,y)}{\partial y}\right|\right]^\gamma\right\}^{\frac{1}{\gamma}}} \tag{S8}$$

where $\gamma$ is a parameter of the model describing the softness of the crossover between low-field and high-field mobilities, and $\mu_{\text{LF}}$ refers to the low-field carrier mobility. Saturation velocity $v_{\text{sat}}$ is related to optical phonon emission energy $\hbar\Omega$, the carrier concentration and the temperature by the following equation:

$$v_{\text{sat}}(y) = \frac{2\Omega}{\pi\sqrt{\pi\rho_{\text{sh}}(y)}}\sqrt{1-\frac{\Omega^2}{4\pi v_{\text{F}}^2\rho_{\text{sh}}(y)}}\,\frac{1}{N_{\text{OP}}+1} \tag{S9}$$

where $\rho_{sh}(y) = n(y) + p(y)$ is the local carrier concentration. Phonon occupation $N_{OP}$ depends on temperature as:

$$N_{OP} = \frac{1}{e^{\frac{\hbar\Omega}{kT}} - 1} \tag{S10}$$

In summary, given the set of material properties and the dimensions of the GFET, and after selecting a bias point ($V'_{gs}$ and $V'_{ds}$), the simulator solves in a self-consistent way the drift-diffusion transport equation (S2) coupled with the 2D Poisson's equation (S1) together with self-heating equation (S3). The simulator then obtains the stationary values of $I_{ds}$, $T$, $n(y)$, $p(y)$, $\psi(x,y)$ and $V(y)$ as the outputs. In this work, we have used the values presented in Table S1 for the material properties and dimensions of the GFET, which correspond to the fabricated GFET described in the main text.

The extrinsic voltages $V_{gs}$ and $V_{ds}$, connected to the terminals of the GFET, are related to the intrinsic voltages at the active area of the device by the following equations:

$$V_{gs} = V'_{gs} + R_s I_{ds} \tag{S11a}$$

$$V_{ds} = V'_{ds} + (R_d + R_s)I_{ds} \tag{S11b}$$

where $R_s = R_d = R_c/2$ are the series resistances at source and drain, which account for the metal-graphene contact resistance together with the access resistance due to the ungated graphene channel.

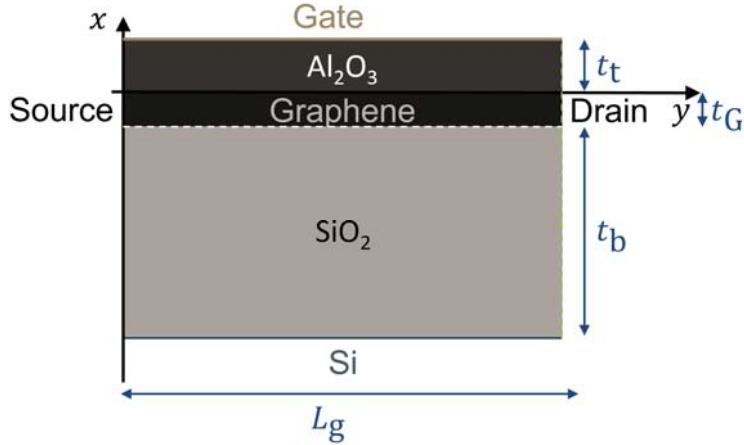

**Figure S1** Cross section of the GFET along channel length and the domain where the Poisson's equation is evaluated. This GFET active area corresponds to the dashed rectangle in figure 1(b) of the main text.

| Parameter | Value |
|:---:|:---:|
| $t_t$ | 22 nm |
| $t_b$ | 1 μm |
| $\varepsilon_t$ | 7.5 $\varepsilon_0$ |
| $\varepsilon_b$ | 3.9 $\varepsilon_0$ |
| $t_G$ | 3.3 $\varepsilon_0$ |
| $L_g$ | 500 nm |
| $W_g$ | 2×15 μm |

**Table S1** Parameters of the simulated GFET.

### S3. Definition of diffusion velocity and total velocity

In the drift-diffusion current given by Eq. (S2), we can define the total carrier velocity, $v_{\text{tot}}$, as the ratio between $I_{\text{ds}}$ and the total carrier concentration, so:

$$v_{\text{tot}}(y) = -\frac{I_{\text{ds}}}{q[n(y)+p(y)]} = -\mu(y)\frac{dV(y)}{dy} \qquad (S12)$$

We can now separate the current into its drift and diffusion contributions. Particularly, the drift current can be expressed in terms of the drift velocity, $v_{\text{drift}}$, which is proportional to the electric field $\xi(y)$ which, in turn, can be written in terms of the (negative) gradient of the electrostatic potential $\psi(y)$:

$$v_{\text{drift}}(y) = \mu(y)\xi(y) = -\mu(y)\frac{d\psi(0,y)}{dy} \qquad (S13)$$

Now we can define a diffusion velocity, $v_{\text{diff}}$, related with the diffusion mechanism transport, as the difference between $v_{\text{tot}}$ and $v_{\text{drift}}$.

$$v_{\text{diff}}(y) = v_{\text{tot}}(y) - v_{\text{drift}}(y) = \mu(y)\frac{d}{dy}[\psi(0,y) - V(y)] \qquad (S14)$$

where $\psi(0,y) - V(y)$ corresponds to the local chemical potential. Since carrier velocities $v_{\text{tot}}$, $v_{\text{drift}}$ and $v_{\text{diff}}$, together with $v_{\text{sat}}$, are magnitudes that are defined locally inside the graphene, we can average their values along the channel length in order to study its behavior as a function of the bias. The averaged values are given by the following formulas, and have been represented in figures 8(e) and (f) in the main text.

$$\langle v_{\text{drift}}\rangle = \frac{1}{L_{\text{g}}}\int_0^{L_{\text{g}}} v_{\text{drift}}(y)dy \qquad (S15a)$$

$$\langle v_{\text{diff}}\rangle = \frac{1}{L_{\text{g}}}\int_0^{L_{\text{g}}} v_{\text{diff}}(y)dy \qquad (S15b)$$

$$\langle v_{\text{tot}}\rangle = \frac{1}{L_{\text{g}}}\int_0^{L_{\text{g}}} v_{\text{tot}}(y)dy \qquad (S15c)$$

$$\langle v_{\text{sat}}\rangle = \frac{1}{L_{\text{g}}}\int_0^{L_{\text{g}}} v_{\text{sat}}(y)dy \qquad (S15d)$$

## S4. Transfer characteristics

Figure S2 shows the experimental transfer characteristic of the GFET (symbols). A fitting of the hole branch at low $V_{ds}$ (represented by solid line) has been gotten by using a flatband voltage $V_{gs0}$ = 2.19 V, puddle concentration $\rho_0$ = 2.93·$10^{11}$ cm$^{-2}$, low-field mobility $\mu_{LF}$ = 1970 cm$^2$ V$^{-1}$ s$^{-1}$, and contact resistance $R_c$ = 11 Ω. The asymmetry can be explained by the difference in mobilities of electrons and holes and/or the difference in contact resistances due to the formation of the p-n junction in an ungated region.

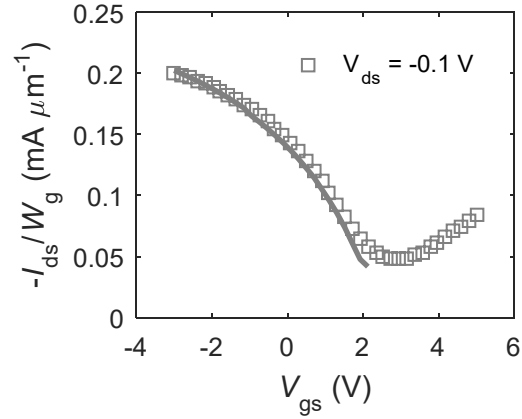

**Figure S2** Measured and simulated transfer curves for the GFET described in the main text.

### S5. Small-signal parameters determination

This section explains how the small-signal matrix $\boldsymbol{Y}(\omega)$ of the GFET is obtained from the stationary model explained in section S2. We consider here the two-port network in common-source configuration represented in figure 3(a) of the main text and we assume that the back gate has a negligible influence over the graphene charge given that the back gate capacitance is much smaller than the top gate capacitance. The charge at the gate, source and drain terminals ($Q_g$, $Q_s$ and $Q_d$, respectively) can be obtained after the evaluation of the charge carrier distribution $q[p(y) - n(y)]$. Upon application of a Ward-Dutton's linear charge partition scheme as the charge control model [9], the terminal charges read as:

$$Q_d = qW_g \int_0^{L_g} \frac{y}{L}[p(y) - n(y)]dy \tag{S16a}$$

$$Q_s = qW_g \int_0^{L_g} \left(1 - \frac{y}{L}\right)[p(y) - n(y)]dy \tag{S16b}$$

$$Q_g = -qW_g \int_0^{L_g}[p(y) - n(y)]dy \tag{S16c}$$

Notice that the total charge in the device is zero, so the model is charge-conserving. From the charge model described above, the intrinsic capacitances of the equivalent circuit shown in figure 3(b) of the main manuscript, can be determined in the following way:

$$C_{gg} = \left.\frac{\partial Q_g}{\partial V\prime_{gs}}\right|_{V\prime_{ds}} \tag{S17a}$$

$$C_{gd} = -\left.\frac{\partial Q_g}{\partial V\prime_{ds}}\right|_{V\prime_{gs}} \tag{S17b}$$

$$C_{dg} = -\left.\frac{\partial Q_d}{\partial V\prime_{gs}}\right|_{V\prime_{ds}} \tag{S17c}$$

$$C_{dd} = \left.\frac{\partial Q_d}{\partial V\prime_{ds}}\right|_{V\prime_{gs}} \tag{S17d}$$

$$C_{gs} = C_{gg} - C_{gd} \tag{S17e}$$

$$C_{sd} = C_{dd} - C_{gd} \tag{S17d}$$

To complete the small-signal model, the transconductance $g_m$ and output conductance $g_{sd}$ need to be evaluated:

$$g_m = \left.\frac{\partial I_{ds}}{\partial V\prime_{gs}}\right|_{V\prime_{ds}} \tag{S18a}$$

$$g_{sd} = \left.\frac{\partial I_{ds}}{\partial V\prime_{ds}}\right|_{V\prime_{gs}} \tag{S18b}$$

As can be deduced from the diagram depicted in figure 3(b), the intrinsic admittance matrix then takes the form:

$$Y'(\omega) = \begin{bmatrix} j\omega C_{\text{gg}} & -j\omega C_{\text{gd}} \\ g_{\text{m}} - j\omega C_{\text{dg}} & g_{\text{sd}} + j\omega C_{\text{dd}} \end{bmatrix} \tag{S19}$$

We must include the influence of the parasitic series resistances $R_{\text{g}}$, $R_s$ and $R_{\text{d}}$ (where $R_{\text{g}}$ is the series resistance at the gate) and parasitic capacitances at the input $C_{\text{pgs}}$ and output $C_{\text{pds}}$ of the two-port network, as can be observed in figure 3(c) of the main text, to obtain the extrinsic admittance matrix $Y(\omega)$. Then, we define the series resistance matrix $Z_{\text{c}}$ and the parasitic capacitance matrix $C_{\text{p}}$ as:

$$Z_{\text{c}} = \begin{bmatrix} R_{\text{g}} + R_{\text{s}} & R_{\text{s}} \\ R_{\text{s}} & R_{\text{d}} + R_{\text{s}} \end{bmatrix} \tag{S20a}$$

$$C_{\text{p}} = \begin{bmatrix} C_{\text{pgs}} & 0 \\ 0 & C_{\text{pds}} \end{bmatrix} \tag{S20b}$$

The extrinsic admittance matrix is calculated from the intrinsic one adding the effect of series resistances and parasitic capacitances as follows:

$$Y(\omega) = \{[Y'(\omega)]^{-1} + Z_{\text{c}}\}^{-1} + j\omega C_{\text{p}} \tag{S21}$$

## S6. Effect of interface trap density

In this section we have analyzed the effect of the interface trap density on the DC and RF behavior. We have simulated the GFET with the parameters given in Table 1 of the main text with $N_{it}$ = 0, $10^{12}$ and $10^{13}$ eV$^{-1}$ cm$^{-2}$. Figure S3 shows that the current-voltage curves simulated with $N_{it}$ = $10^{12}$ eV$^{-1}$ cm$^{-2}$ do not differ significantly from the case without traps. The situation strongly changes for $N_{it}$ = $10^{13}$ eV$^{-1}$ cm$^{-2}$, where the high amount of charged defects clearly makes the carrier concentration at a given bias to decrease, which reduces the mobile charge and, thus, the total drain current. However, the highest possible $f_{T,x}$ and $f_{max}$ that can be achieved considering any of the three examined $N_{it}$ are quite similar. Figure S4 presents the RF figures of merit as a function of the bias point, using the values of the parasitic elements of Table 2 in the main text. The maxima of $f_{T,x}$ and $f_{max}$ reach approximately 25 and 40 GHz independently of $N_{it}$ although they are located at different biases: as the density of defects grows, maxima move off from the Dirac voltage. It can thus be concluded that, in our model, $N_{it}$ up to a level of $10^{12}$ eV$^{-1}$ cm$^{-2}$, does not influence the best RF performance of the GFET but it affects the bias that optimize them.

Notice that charged traps are assumed here to not change with the rapid variations of the small-signal voltage [10]. That is, the mean time of trapping and detrapping charges are larger than the period of the RF signal. In case that charged defects were affected by the small-signal voltage, RF performance would decrease considerably.

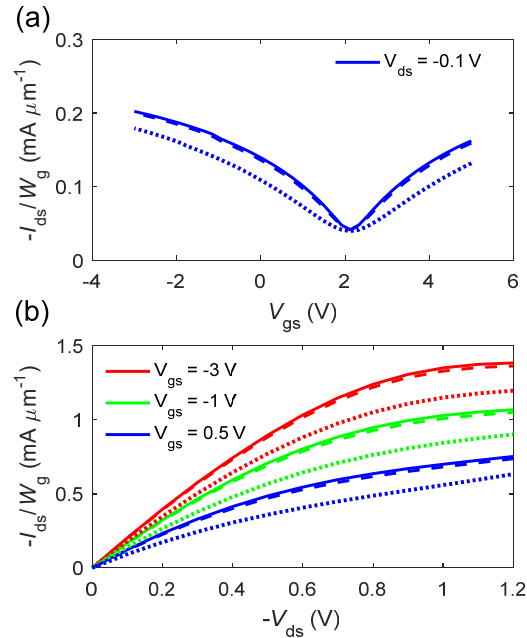

**Figure S3** Influence of interface trap density on the (a) transfer and (b) output curves. Solid lines correspond to $N_{it}$ = 0 eV$^{-1}$ cm$^{-2}$; dashed lines to $N_{it}$ = $10^{12}$ eV$^{-1}$ cm$^{-2}$; and dotted lines to $N_{it}$ = $10^{13}$ eV$^{-1}$ cm$^{-2}$.

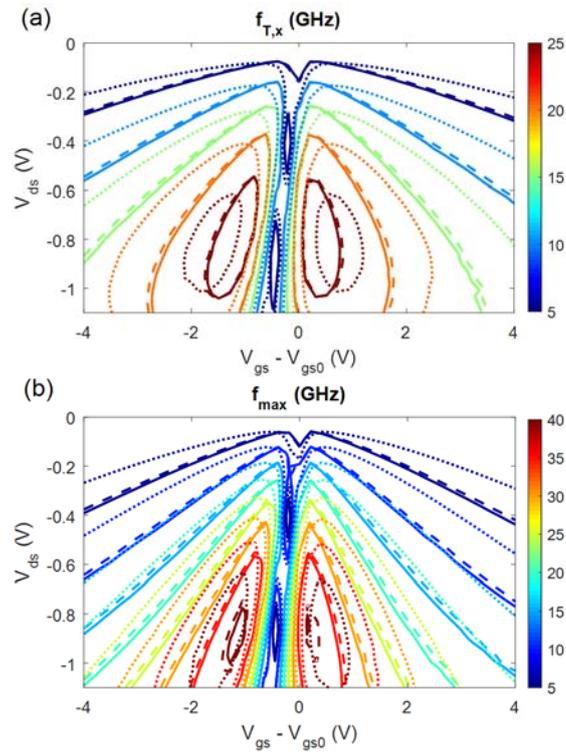

**Figure S4** Influence of the interface trap density on RF figures of merit: (a) $f_{T,x}$ and (b) $f_{max}$. Solid lines correspond to a $N_{it} = 0$ eV$^{-1}$ cm$^{-2}$; dashed lines to $N_{it} = 10^{12}$ eV$^{-1}$ cm$^{-2}$; and dotted lines to $N_{it} = 10^{13}$ eV$^{-1}$ cm$^{-2}$.

## S7. Effect of self-heating on $f_{max}$

We have compared the largest $f_{max}$ that can be achieved with and without accounting the self-heating phenomena. The latter assumes that the thermal resistance is null, so graphene channel remains at room temperature. Figure S5 shows the bias dependence of $f_{max}$ in both cases, showing that self-heating severely limit RF performance of GFETs. Specifically, $f_{max}$ over 60 GHz can be reached for the studied bias window considering that the device operates at room temperature.

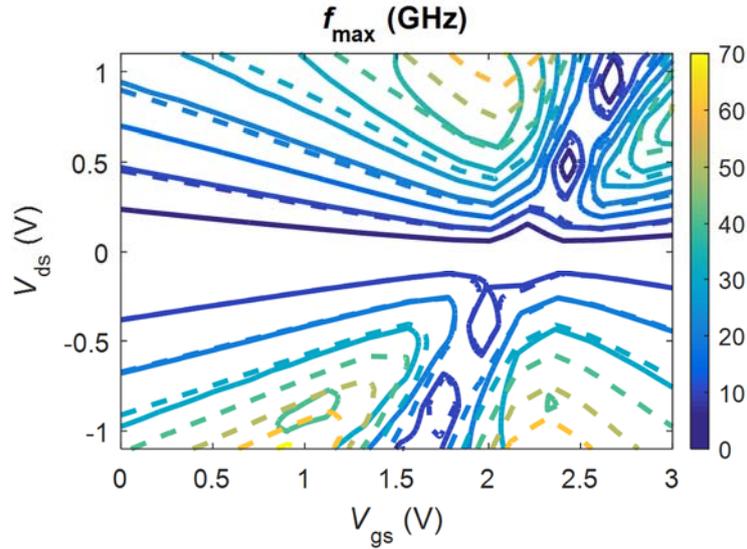

**Figure S5** Map of $f_{max}$ as a function of the bias point for GFET including the self-heating effect (solid lines) and switching off such a phenomena (dashed lines).